\documentclass[12pt]{article}
\usepackage{graphicx}
\usepackage[percent]{overpic}

\newcommand\pubnumber{PNNL-SA-106625}
\newcommand\pubdate{\today}

\def\pnnl{Pacific Northwest National Laboratory\\
902 Battelle Boulevard, 99352 - Richland, WA, USA,\\
for the Belle and BaBar Collaborations.}

\def\Title#1{\begin{center} {\Large #1 } \end{center}}
\def\Author#1{\begin{center}{ \sc #1} \end{center}}
\def\Address#1{\begin{center}{ \it #1} \end{center}}

\newcommand\pubblock{\rightline{\begin{tabular}{l} \pubnumber\\
         \pubdate  \end{tabular}}}
\newenvironment{Abstract}{\begin{quotation}  }{\end{quotation}}
\newenvironment{Presented}{\begin{quotation} \begin{center} 
             PRESENTED AT\end{center}\bigskip 
      \begin{center}\begin{large}}{\end{large}\end{center} \end{quotation}}

\def\beq{\begin{equation}}
\def\eeq#1{\label{#1}\end{equation}}
\def\eeqn{\end{equation}}

\def\beqa{\begin{eqnarray}}
\def\eeqa#1{\label{#1}\end{eqnarray}}
\def\eeqan{\end{eqnarray}}

\let\bar=\overbar

\def\Dslash{\not{\hbox{\kern-4pt $D$}}}
\def\dslash{\not{\hbox{\kern-2pt $\del$}}}

\def\msb{{\bar{\ssstyle M \kern -1pt S}}}

\def\babar{\mbox{\slshape B\kern-0.1em{\small A}\kern-0.1em
    B\kern-0.1em{\small A\kern-0.2em R}}}
\def\babartbl{\mbox{\slshape B\kern-0.1em{\scriptsize A}\kern-0.1em
    B\kern-0.1em{\scriptsize A\kern-0.2em R}}}
\def\fours{\ensuremath{\Upsilon(4S)}}

\def\beq{\begin{equation}}
\def\eeq{\end{equation}}

\def\BtoDsttaunuGen{\ensuremath{B\to \bar D^{(*)} \tau^+ \nu_\tau}}

\def\BztoDsttaunu{\ensuremath{B^0\to D^{*-} \tau^+ \nu_\tau}}
\def\BztoDtaunu{\ensuremath{B^0\to D^{-} \tau^+ \nu_\tau}}
\def\BptoDsttaunu{\ensuremath{B^+\to \bar D^{*0} \tau^+ \nu_\tau}}
\def\BptoDtaunu{\ensuremath{B^+\to \bar D^{0} \tau^+ \nu_\tau}}

\def\BtoDstlnuGen{\ensuremath{B\to \bar D^{(*)} \ell^+ \nu_\ell}}

\def\RDstGen{\ensuremath{R(D^{(*)})}}
\def\RD{\ensuremath{R(D)}}
\def\RDst{\ensuremath{R(D^*)}}

\newcommand{\tev}{\ensuremath{\mathrm{Te\kern -0.1em V}}}
\newcommand{\gev}{\ensuremath{\mathrm{Ge\kern -0.1em V}}}	
\newcommand{\mev}{\ensuremath{\mathrm{Me\kern -0.1em V}}}	
\newcommand{\kev}{\ensuremath{\mathrm{ke\kern -0.1em V}}}	

\begin{document}
\begin{titlepage}
\pubblock

\vfill
\Title{New physics searches in $B \rightarrow D^{(*)}\tau \nu$ decays }
\vfill
\Author{ Vikas Bansal}
\Address{\pnnl}
\vfill
\begin{Abstract}
I review the current status of measurements involving semi-tauonic  $B$ meson decay at the $B$-factories. I briefly discuss the experimental methods and highlight the importance of background contributions especially from poorly understood $D^{**}$ in this study. Perhaps this can also shed some light on the discrepancy in the BaBar measurement of ratio of semi-tauonic and semi-leptonic ($e$/$\mu$) modes of $B$ decay from the Standard Model (SM) at 3.2$\sigma$. I will also discuss one of the New Physics (NP) models that could be experimentally sensitive in being distinguished from the Standard Model (SM).
\end{Abstract}
\vfill
\begin{Presented}
 8th International Workshop on the CKM Unitarity Triangle (CKM 2014), Vienna, Austria, September 8-12, 2014.
\end{Presented}
\vfill
\end{titlepage}
\def\thefootnote{\fnsymbol{footnote}}
\setcounter{footnote}{0}

\section{Introduction}

Search for New Physics (NP) via $b \rightarrow c \tau \nu_{\tau}$ transitions is particularly interesting as it involves third-generation fermions both in the initial and final states. Presence of leptoquarks and charged Higgs boson in 2 Higgs doublet model (2HDM) could enhance or suppress this decay rate~\cite{tanaka2013}. In addition, study of $\tau$ polarizations in its hadronic decay modes can also be sensitive to NP~\cite{fajfer2012}.

Reconstruction of $B$ meson exclusive decay with $\tau$ lepton leads to two or three undetected neutrinos depending on the $\tau$ decay mode. This requires additional constraints related to $B$ meson production as are available at the $B$-factory experiments - BaBar~\cite{Aubert:2001tu} and Belle~\cite{Abashian:2000cg}. $B$-factories produce \fours\ that almost exclusively decays into $B\bar{B}$ system, thereby allowing full event reconstruction which is not possible at the LHCb~\cite{LHCb:2008}. {\it Hadronic tagging} and {\it inclusive tagging} are two such popular reconstruction techniques.  

\section{Experimental Results}

\begin{figure}[h]
\centering
\includegraphics[width=3.5in]{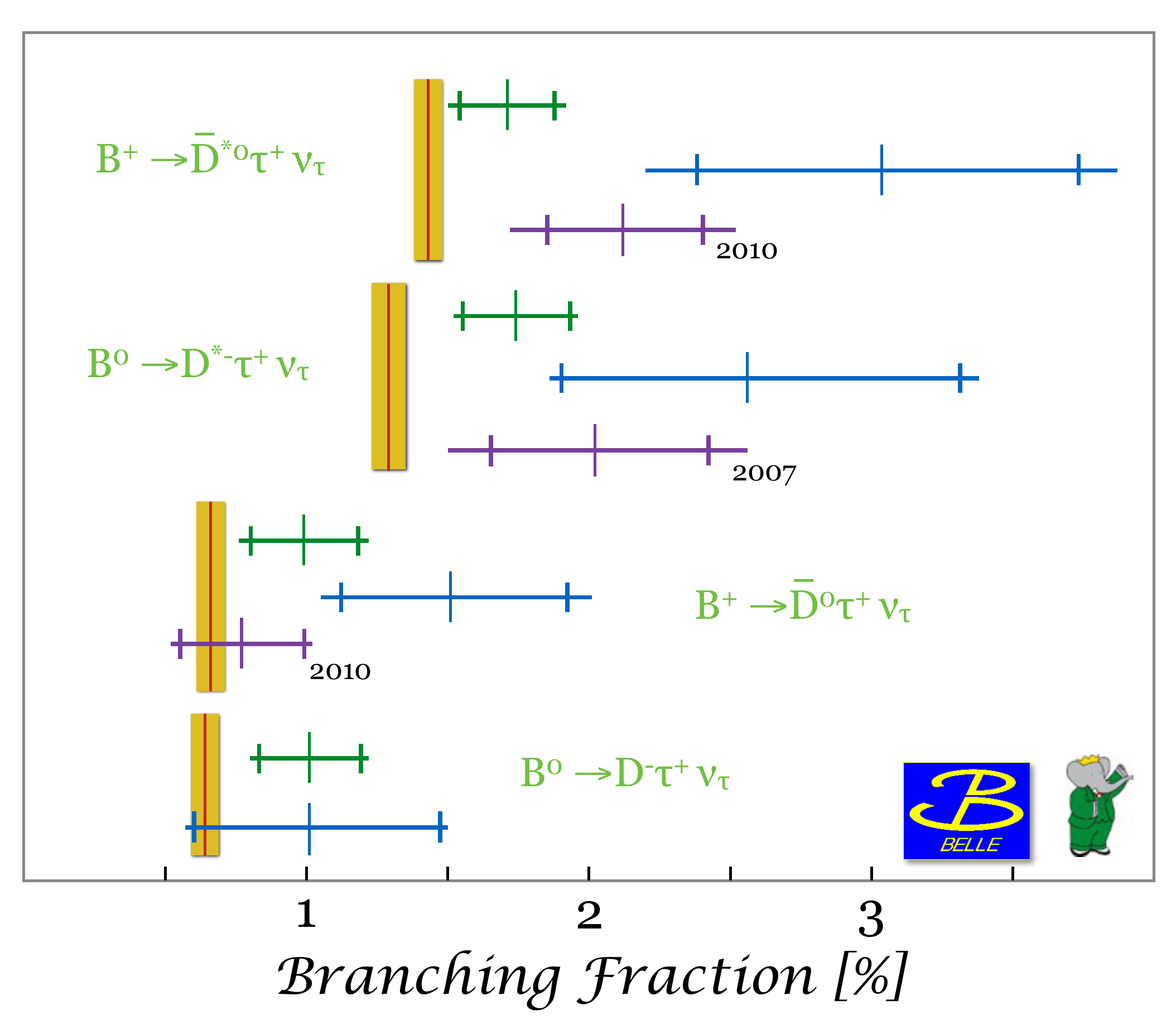}
\caption{All four $B \rightarrow D^{(*)}\tau \nu$ branching fractions. Red vertical line shows the standard model prediction with $\pm 1\sigma$ uncertainty yellow bands. Horizontal lines span the measurement by $\pm 1\sigma$ total uncertainty where blue and purple are for Belle and green is for BaBar. One larger vertical line and two smaller thicker vertical lines cross each horizontal line. Larger line denotes the measured central value and the region enclosed by the two smaller lines includes $\pm 1\sigma$ statistical uncertainty only.}
\label{fig:Dtaunu_BF}
\end{figure}

Belle first observed $B \rightarrow D^{(*)}\tau \nu$ decays in 2007 using inclusive tagging method~\cite{Matyja:2007kt}. In 2010, Belle measured the decay rates \BptoDsttaunu\ and \BptoDtaunu\ with
the same analysis technique and a larger data sample and additional $D$ decay modes~\cite{Bozek:2010xy}. BaBar performed a hadronic tagging measurement of all four channels \BztoDtaunu\, \BztoDsttaunu, \BptoDtaunu, and \BptoDsttaunu\ in 2008 which was superseded by their 2012 result~\cite{ref:bbr-dtaunu-2012,ref:bbr-dtaunu-2013} using full BaBar dataset. Belle also used hadronic tagging technique to measure the four branching fractions~\cite{Adachi:2009qg}.  All results along with the standard model (SM) prediction~\cite{fajfer2012} are summarized in Fig.~\ref{fig:Dtaunu_BF}, where purple, blue and green lines denote Belle inclusive tagging, Belle hadronic tagging and BaBar hadronic tagging results, respectively, and red lines with yellow bands show SM prediction.

Many experimental and theoretical uncertainties cancel out or are reduced when one measures the ratios of the decay rates,
\beq
\RDstGen \equiv {\Gamma(\BtoDsttaunuGen)\over \Gamma(\BtoDstlnuGen)}.
\eeq 

\begin{figure}[t]
\centering
\includegraphics[width=3.5in]{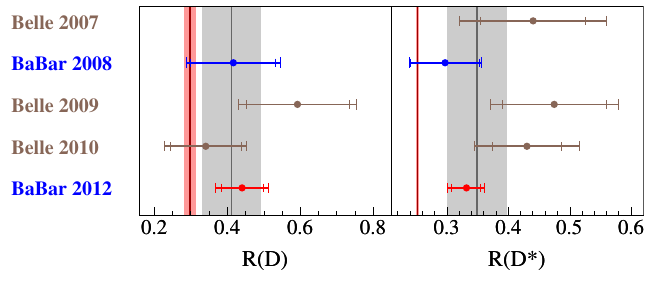}
\caption{\RD\ and \RDst\ measurements with statistical and total uncertainties. The vertical bands represent the average of the 2007-2010 measurements (light shading) and SM predictions (dark shading) for both ratios separately. The widths of the bands represent the uncertainties.}
\label{fig:RDst}
\end{figure}

Figure~\ref{fig:RDst} summarizes the measurements of the ratios \RD\ and \RDst\ along with the SM prediction. Combination of the ratios gives an observed excess over the SM by 3.2$\sigma$ significance~\cite{ref:bbr-dtaunu-2013}.

\begin{figure}[htb]
\centering
\includegraphics[width=3.5in]{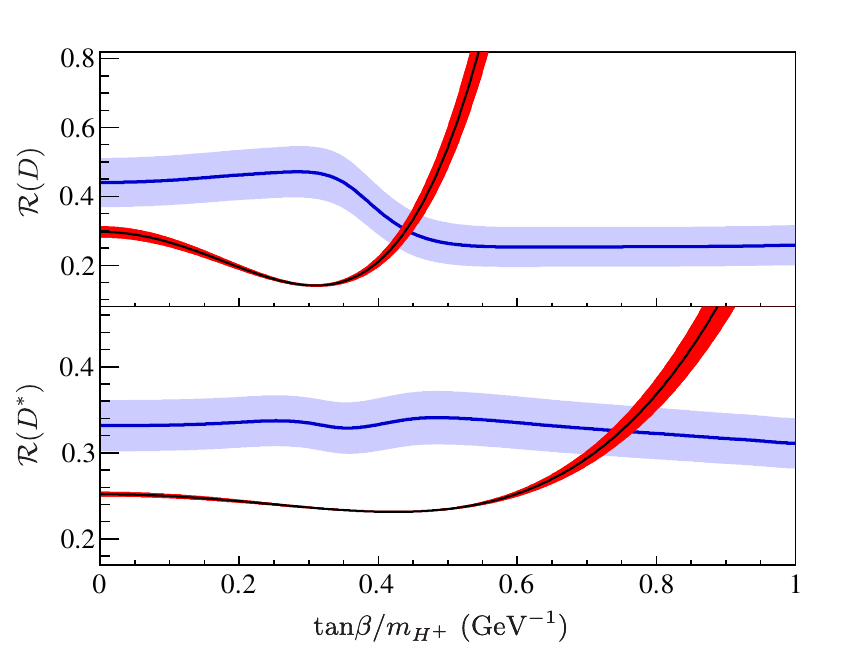}
\caption{Comparison of the BaBar results~\cite{ref:bbr-dtaunu-2012} (light band, blue)  with predictions that include charged Higgs boson of type II 2HDM (dark band, red). The widths of the two bands represent the uncertainties. The SM corresponds to $\tan\beta/m_{H^+} = 0$.}
\label{fig:chHiggEx}
\end{figure}

Figure~\ref{fig:chHiggEx} compares the measured values of \RD\ and \RDst\ \cite{ref:bbr-dtaunu-2012} in the context of the type-II 2HDM to the theoretical predictions as a function of $\tan\beta/m_{H^+}$. While the measured values match the predictions $\tan\beta/m_{H^+} = 0.44 \pm 0.02~\gev^{-1}$ from \RD\ and $\tan\beta/m_{H^+} = 0.75 \pm 0.04~\gev^{-1}$ from \RDst, the combination of the two is excluded at $99.8\%$ confidence level for any value of $\tan\beta/m_{H^+}$.

In $b \rightarrow c \tau \nu_{\tau}$ transitions, measured exclusive rates do not sum up to the measured inclusive rate \ensuremath{B\to X_{c} \ell \nu_\ell}~\cite{hfag} in these decay modes. One of the reasons is many unobserved $D^{**}$ states in these modes~\cite{bigi:2007}. These states are also a major source of uncertainties in any measurement of \RD, \RDst\ or the  $B \rightarrow D^{(*)}\tau \nu$ branching fractions. BaBar measurement~\cite{ref:bbr-dtaunu-2013} for ratios \RDstGen\ makes a good attempt in assessing $D^{**}$ contribution. It takes in to account $1P$ excited $D^{**}$ states and estimates  $D^{**}$ contribution in $\tau$ channel from light lepton $\ell$ channel by using available phase space. But these type of studies are still vulnerable to poorly understood $D^{**}$.

\section{New Physics Models}

\begin{figure}[htb]
\centering
\includegraphics[width=3.5in]{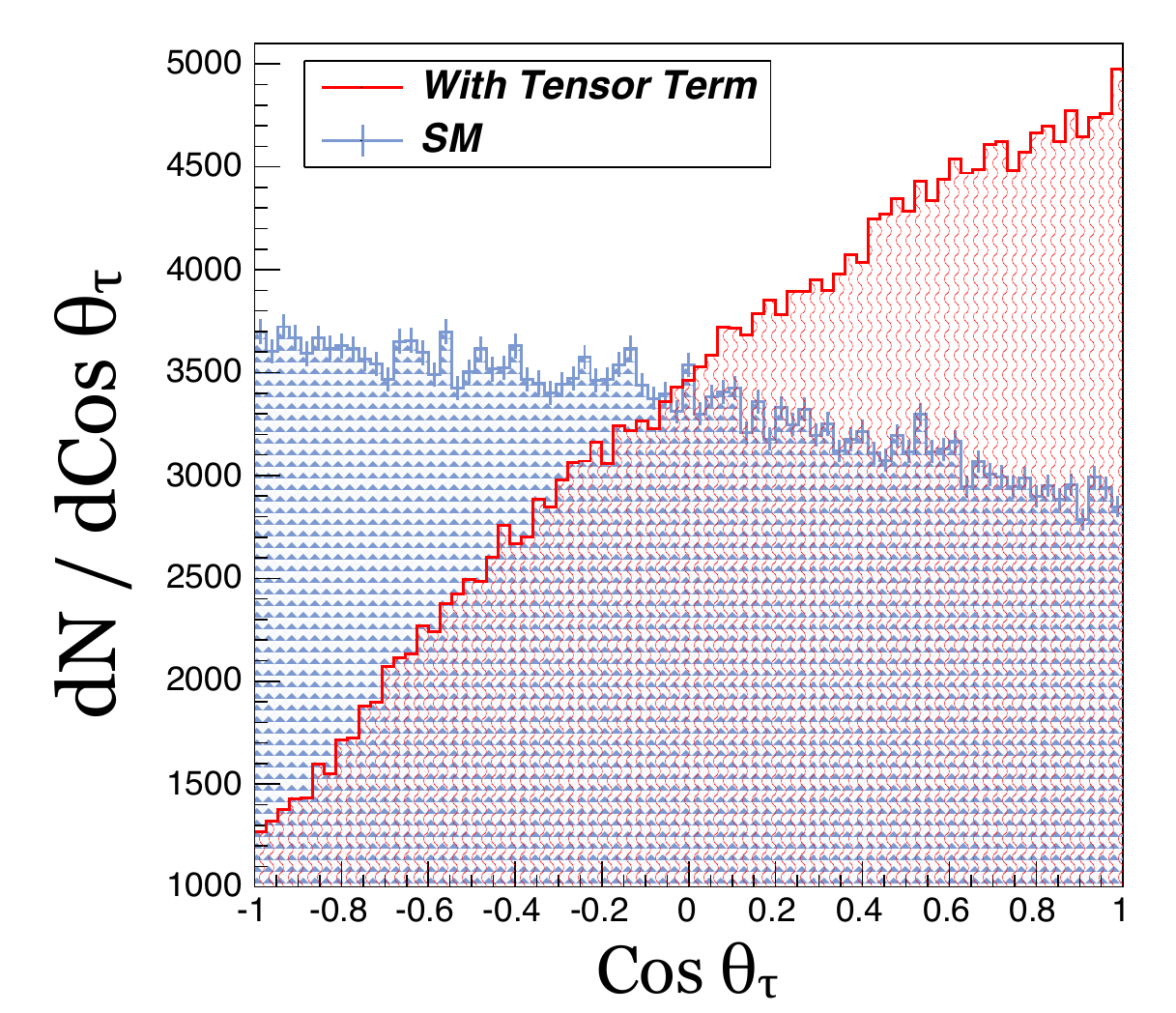}
\caption{Differential Cos $\theta_{\tau}$ distribution for the tensor operator~\cite{tanaka2013} (red) and the SM (blue).}
\label{fig:newModel}
\end{figure}

Figure~\ref{fig:newModel} compares differential distribution of Cos $\theta_\tau$ for the tensor operator in a model-independent approach and the SM following this methodology in Ref.~\cite{tanaka2013}, where $\theta_{\tau}$ is the angle between the momentum vector of the tau lepton and the mediating meson in the rest frame of the meson. The angle is sensitive in distinguishing model parameters from the SM. From an experimental perspective, it will be useful to know which few NP points could be simulated so as to scan a wide range of parameter space by interpolation.

\end{document}